\documentclass{article}
\usepackage{spconf,amsmath,graphicx}

\usepackage{enumitem}
\setlist{nosep, leftmargin=14pt}

\usepackage{mwe} % to get dummy images
\usepackage{amssymb}
\usepackage{algorithm}
\usepackage{algpseudocode}

\usepackage{microtype}
\usepackage{hyperref}
\usepackage{url}
\usepackage{paralist}
\usepackage{colortbl}
\usepackage{booktabs}
\usepackage{amsmath}
\usepackage{lineno}
\usepackage{graphicx}
\usepackage{subfigure}
\usepackage{float}
\usepackage{caption}
\usepackage{wrapfig}
\usepackage{adjustbox}

\newcommand{\foo}{\textcolor{blue}{$\bullet$} \hspace{5pt}}

\title{\texttt{RAA}-MIL: A Novel Framework for Classification of Oral Cytology}
\name{
Rupam Mukherjee$^{\dagger1}$\;\;
Rajkumar Daniel$^{\dagger1}$\;\;
Soujanya Hazra$^{\dagger2}$\;\;
Shirin Dasgupta$^{3}$\;\;
Subhamoy Mandal$^{1}$
\thanks{$^{\dagger}$These authors contributed equally.}
}

\address{
$^{1}$ School of Medical Science \& Technology, IIT Kharagpur, India\\
$^{2}$ Department of Electrical Engineering, IIT Kharagpur, India\\
$^{3}$ Dr. B.C. Roy Multispeciality Medical Research Centre, IIT Kharagpur, India
}

\begin{document}
%\ninept
%
\maketitle
\begin{abstract}
Cytology is a valuable tool for early detection of oral squamous cell carcinoma (OSCC). However, manual examination of cytology whole slide images (WSIs) is slow, subjective, and depends heavily on expert pathologists. %To reduce time and inter-observer variation, automated analysis methods are required. Such development relies on high-quality, multi-centre datasets. One significant step in this direction is the recently released \textit{Oral Cytology Dataset} \cite{Jain2025Cytology}, which provides annotated cytology WSIs from ten medical centres across India. Building upon this resource, we present the first weakly supervised deep learning framework for patient-level diagnosis of oral cytology slides.
To address this, we introduce the first weakly supervised deep learning framework for patient-level diagnosis of oral cytology whole slide images, leveraging the newly released \textit{Oral Cytology Dataset}~\cite{Jain2025Cytology}, which provides annotated cytology WSIs from ten medical centres across India.  Each patient case is represented as a bag of cytology patches and assigned a diagnosis label (\textit{Healthy}, \textit{Benign}, \textit{Oral Potentially Malignant Disorders (OPMD)}, \textit{OSCC}) by an in-house expert pathologist. These patient-level weak labels form a new extension to the dataset. We evaluate a baseline multiple-instance learning (MIL) model and a proposed \textbf{Region-Affinity Attention MIL (\texttt{RAA}-MIL)} that models spatial relationships between regions within each slide. The \texttt{RAA}-MIL achieves an average accuracy of \textbf{72.7\%}, weighted F1-score of \textbf{0.69} on an unseen test set, outperforming the baseline. This study establishes the first patient-level weakly supervised benchmark for oral cytology and moves toward reliable AI-assisted digital pathology.
\end{abstract}

\begin{keywords}
Oral Cytology, Weak Supervision, Patient-Level Classification, Multiple Instance Learning, Region-Affinity Attention, Deep Learning, Oral Cancer, Digital Pathology
\end{keywords}

\section{Introduction}

OSCC is a major public health concern and one of the most common head and neck malignancies. Early detection greatly improves survival outcomes, yet diagnosis often occurs at advanced stages. Histopathology remains the diagnostic gold standard but is invasive, labor-intensive, and depends on expert pathologists. Oral exfoliative cytology offers a simpler adjunct technique for screening, but manual evaluation of WSIs is slow, subjective, and prone to inter-observer variability. These limitations motivate automated, reliable computational methods for cytological diagnosis.

Deep learning has rapidly advanced computational cytology. Jiang \textit{et al.}~\cite{JIANG2023102691} reviewed over a hundred studies and highlighted the shift toward weakly supervised and transformer-based methods for slide-level prediction. Sukegawa \textit{et al.}~\cite{Sukegawa2024} demonstrated that diverse expert annotations improve CNN-based oral cytology classification. Wang and Zheng~\cite{Wang2024OralCytology} noted the growing role of AI-assisted analysis in oral exfoliative cytology. Beyond classification, transformer-based affinity modeling~\cite{Ru2022LearningAffinity} has shown promise in capturing contextual relationships among regions, suggesting new directions for cytology slide interpretation.

A recent milestone for this domain is the multi-institutional \textit{Oral Cytology Dataset} by Jain \textit{et al.}~\cite{Jain2025Cytology}. %It includes 368 high-resolution PAP- and MGG-stained WSIs with detailed cell and region-level annotations from ten medical centers across India. 
This dataset enables reproducible AI development for segmentation and cell-level morphology analysis. However, it lacks diagnostic labels at the patient level, which are essential for clinically meaningful classification. In practice, pathologists integrate information across multiple regions before issuing a patient-level diagnosis. Building on this foundation, we introduce the first weakly supervised deep learning framework for patient-level oral cytology diagnosis. Our work makes three key contributions:

\noindent\foo\hspace{-.1cm}\textit{\textbf{Dataset}:} We extend the \textit{Oral Cytology Dataset}~\cite{Jain2025Cytology} by introducing expert-verified, patient-level weak labels across four diagnostic categories: \textit{Healthy}, \textit{Benign}, \textit{OPMD}, and \textit{OSCC}. These labels enable learning at the patient level, aligning with real clinical workflows.
    
\noindent\foo\hspace{-.1cm}\textit{\textbf{Framework}:} We present the first weakly supervised deep learning framework for oral cytology. Each patient is represented as a bag of cytology regions, forming a MIL formulation for diagnosis.  We propose a novel \textbf{Region-Affinity Attention MIL (\texttt{RAA}-MIL)} model that learns relationships between cytological regions within each slide. The affinity mechanism enhances contextual understanding and improves diagnostic accuracy.
    
\noindent\foo\hspace{-.1cm}\textit{\textbf{Benchmark}:} We establish the first patient-level benchmark for automated oral cytology classification. The proposed \texttt{RAA}-MIL achieves a mean accuracy of \textbf{72.7\%} and a weighted F1-score of \textbf{0.69}, outperforming the baseline MIL model.

These contributions collectively move oral cytology toward reproducible, AI-assisted diagnostic workflows that can support faster and more consistent clinical screening.

\section{Framework}

\subsection{Dataset}

This study uses the \textit{Oral Cytology Dataset} introduced by Jain \textit{et al.}~\cite{Jain2025Cytology}. The full dataset contains 368 PAP- and MGG-stained WSIs collected from ten medical centers and digitized at 40× magnification with detailed region and nucleus-level annotations. For this work, only the representative region-level patches provided for each patient were accessible, resulting in a subset of 162 patients. Each patient is treated as a bag, and each bag contains a variable number of high-resolution patches of size 2048\,$\times$\,2048\,px. We extend this subset by assigning expert-verified patient-level weak labels across four diagnostic categories: \textit{Healthy}, \textit{Benign}, \textit{OPMD}, and \textit{OSCC}. This structure enables a MIL formulation for patient-level diagnosis, where the global label is known, but individual patch labels are not.

\subsection{Methodology}
Our goal is to perform patient-level weakly supervised classification of oral cytology cases using only a single diagnosis per patient. Each patient provides a variable number of high-resolution representative patches ($2048\times 2048$), selected by an expert pathologist. These patches are processed as bags in a MIL framework. The pipeline consists of four stages: (1) frozen self-supervised ViT tokenization, (2) region-affinity attention (\texttt{RAA}), (3) gated attention-based MIL pooling, and (4) cross-validated training with prediction-averaging ensemble.

\subsubsection{Patch Tokenization via Frozen SSL ViT}

Each patch is resized to $224\times224$ and embedded using a DINO-pretrained ViT-S/16 encoder as released in the pathology SSL benchmark of Kang \textit{et al.}~\cite{kang2022benchmarking}. 
The encoder remains frozen. Each image is decomposed into $14\times14=196$ non-overlapping tokens, each of dimension $D=384$. The CLS token is discarded following self-supervised pathology conventions.  
For a patient with $P$ patches, we obtain:
\[
X = [\mathbf{x}_1,\ldots,\mathbf{x}_N] \in \mathbb{R}^{N\times D}, \qquad N = 196P.
\]
Tokens are cached to disk for efficiency and determinism during MIL training.

\subsubsection{Region-Affinity Attention (\texttt{RAA})}

ViT patch tokens capture local content but ignore short-range spatial smoothness.  
We introduce \texttt{\texttt{RAA}}, which refines each token through its $k\times k$ neighborhood ($k=3$ in our experiments).

Let $\mathbf{z}_i$ denote the token at position $i$ inside a $14\times14$ grid.  
For each neighbor $j\in\mathcal{N}(i)$, we compute a distance:
\[
d_{ij} = \frac{1}{D}\lVert \mathbf{z}_i - \mathbf{z}_j \rVert_2^2.
\]
A lightweight MLP maps each distance to a scalar affinity, followed by a softmax over the neighborhood:
\[
\alpha_{ij} = \frac{\exp(f(d_{ij}))}{\sum_{\ell\in\mathcal{N}(i)} \exp(f(d_{i\ell}))}.
\]
The refined token is:
\[
\mathbf{z}^{\,\text{out}}_i
= \mathbf{z}_i 
+ \gamma\Big( \mathrm{LN}\big(\sum_{j\in\mathcal{N}(i)} \alpha_{ij}\mathbf{z}_j\big) - \mathbf{z}_i \Big),
\]
where $\gamma$ is a learnable residual gate initialized to zero.  
\texttt{RAA} is applied independently to each patch and preserves the ViT token grid structure.

\subsubsection{Gated Attention MIL Pooling}

After \texttt{RAA} refinement, all tokens are treated as MIL instances.  
Let $\tilde{X} \in \mathbb{R}^{M\times D}$ denote the set of refined tokens ($M = 196P$).  
We adopt the gated attention-based MIL mechanism from Ilse \textit{et al.}~\cite{ilse2018attention}.  
Two parallel transformations produce complementary attention signals:
\[
\mathbf{a} = \tanh(W_a \tilde{X}^\top),\qquad
\mathbf{b} = \sigma(W_b \tilde{X}^\top).
\]
The attention logits and subsequent instance weights are obtained via softmax are:
\[
\mathbf{u} = W_c^\top(\mathbf{a} \odot \mathbf{b}), \quad w_i = \frac{\exp(u_i)}{\sum_j \exp(u_j)}
\]

The bag embedding is:
\[
\mathbf{m} = \sum_{i=1}^M w_i \tilde{\mathbf{x}}_i.
\]
A small MLP classifier $\Phi$ maps $\mathbf{m}$ to patient-level logits.

\subsubsection{Training and Optimization}

Only the \texttt{RAA} module and MIL classifier are trained.  
The ViT encoder remains frozen.  
We use class-weighted focal loss with label smoothing, AdamW optimizer, ReduceLROnPlateau scheduling on validation F1, and early stopping.  
Training is performed with 5-fold stratified cross-validation on patients, with batch size one bag per step.

\subsubsection{Prediction-Averaging Ensemble}

\begin{figure*}[htb]
\centering
\includegraphics[width=\textwidth]{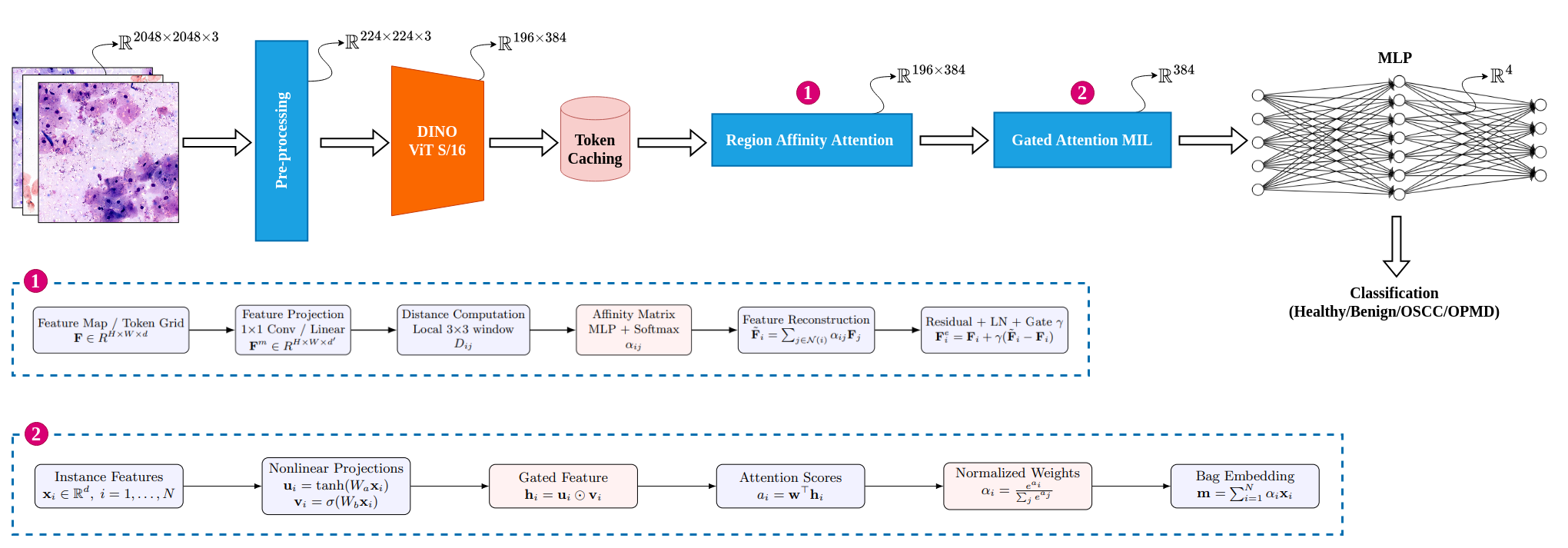}
\caption{
Overview of the proposed \texttt{RAA}–MIL pipeline. Frozen ViT tokens are refined using \texttt{RAA} and pooled with gated MIL to produce a patient-level embedding, which a lightweight MLP maps to the final four-class diagnosis.
}
\label{fig:architecture}
\end{figure*}

For final evaluation, we load the best model from each fold and perform inference on the held-out test subset.  
For test bag $i$, fold $f$ produces probability vector $\mathbf{p}_i^{(f)}$.  
The ensemble output is the arithmetic mean:
\[
\bar{\mathbf{p}}_i = \frac{1}{F} \sum_{f=1}^{F} \mathbf{p}_i^{(f)},
\qquad
\hat{y}_i = \arg\max_c \bar{p}_{i,c}.
\]
This reduces MIL-induced variance and improves stability.  
Figure~\ref{fig:architecture} summarizes the full forward pass.

\subsubsection{Experimental Setup}

We evaluated both models under stratified five-fold cross-validation
(Stratified 5-fold CV) to ensure class-balanced partitions.
Cross-validation was performed on the training split only. The held-out
test set (\(N=33\) patients) was never used during model development.
This design completely eliminates any possibility of data leakage.
All experiments follow a four-class setting:
\textit{Healthy}, \textit{Benign}, \textit{OPMD},
and \textit{OSCC}. We report accuracy,
weighted F1-score, receiver operating characteristic area under the curve
(ROC-AUC), and precision–recall area under the curve (PR-AUC). Final
test predictions use ensemble averaging across the five CV models.

\section{Results}

\noindent\textbf{Cross-Validation Performance:} Stratified 5-fold CV shows stable training dynamics under weak
supervision. The vanilla Attention-based Multiple Instance Learning
(MIL) baseline  achieved a mean validation accuracy of
\(0.588 \pm 0.059\) and a weighted F1-score of \(0.579 \pm 0.077\).
The proposed Region Affinity Attention MIL (\texttt{RAA}-MIL) improved these
to \(0.619 \pm 0.062\) accuracy and \(0.608 \pm 0.063\) weighted F1.
The low standard deviations across folds confirm stable convergence and
robust MIL pooling behavior. These CV trends suggest that \texttt{RAA}-MIL
learns more discriminative representations under weak patient-level
labels.

\noindent\textbf{Hold-out Test Performance:} Table~\ref{tab:test} summarizes the final ensemble performance on the
test set. \texttt{RAA}-MIL outperforms the baseline in all key metrics.
The ensemble accuracy improved from \(69.7\%\)  to \(72.7\%\) ,
and the weighted F1-score increased from \(0.6749\) to \(0.6970\).
Weighted PR-AUC also increased substantially from \(0.7389\) to
\(0.7969\), reflecting more reliable probability calibration.

\begin{table}[h]
\small
\centering
\caption{Hold-out test performance using ensemble prediction averaging.}
\label{tab:test}
\scalebox{0.85}{  %
\begin{tabular}{lccc}
\hline
\textbf{Model} & \textbf{Accuracy} & \textbf{F1 (Weighted)} & \textbf{AUPRC (Weighted)} \\
\hline
Vanilla MIL  & 0.6970 & 0.6749 & 0.7389 \\
\texttt{RAA}-MIL      & \textbf{0.7273} & \textbf{0.6970} & \textbf{0.7969} \\
\hline
\end{tabular}
}
\end{table}

\begin{figure*}[ht]
\centering
\includegraphics[width=0.9\textwidth]{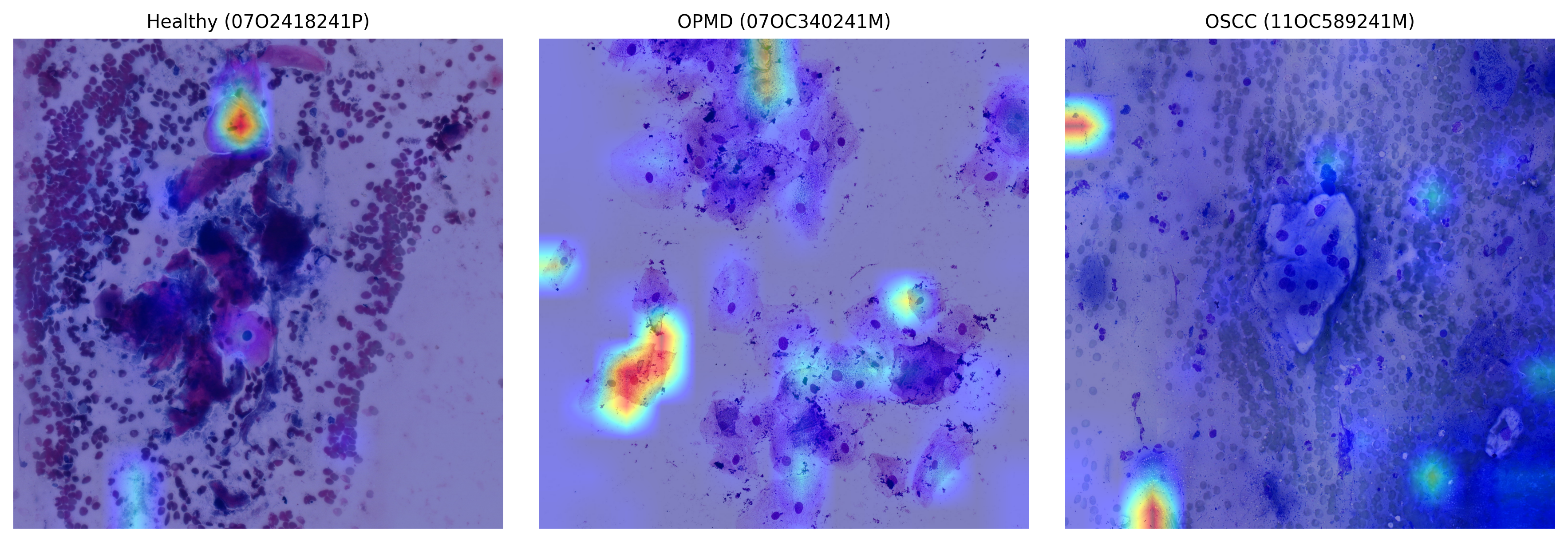}
\caption{
Qualitative visualization of gated-MIL attention maps for
three representative cases from the test set: Healthy, OPMD, and OSCC.
% Warm colors indicate regions with higher attention weights. The model
% successfully localizes diagnostic cellular clusters, with minimal activation
% on background artefacts, demonstrating clinically meaningful feature usage
% under weakly supervised training.
}
\label{fig:attn_maps}
\end{figure*}

\noindent\textbf{Class-wise Observations:} Table~\ref{tab:prauc} reports class-specific PR-AUC values.
\textit{Healthy} achieved the highest PR-AUC (\(0.9441\)), aided by
strong morphological separability and adequate representation.
\textit{OPMD} performance improved significantly in v2
(\(0.7594\)), reflecting better discrimination of premalignant
features. \textit{OSCC} also showed strong improvement
(\(0.7667\)), demonstrating that \texttt{RAA}-MIL captures malignant cytological
patterns despite limited data. The \textit{Benign} class remains
challenging for both models due to extremely low test support
(\(n=3\)). This limitation stems from dataset imbalance rather than model
failures.

\begin{table}[h]
\small
\centering
\caption{Per-class PR-AUC comparison on the test set.}
\label{tab:prauc}
\begin{tabular}{lcccc}
\hline
\textbf{Model} & \textbf{Benign} & \textbf{Healthy} & \textbf{OPMD} & \textbf{OSCC} \\
\hline
Vanilla MIL 
& 0.1879 & 0.9193 & 0.6937 & 0.5437 \\
\texttt{RAA}-MIL 
& 0.1772 & \textbf{0.9441} & \textbf{0.7594} & \textbf{0.7667} \\
\hline
\end{tabular}
\end{table}

\noindent\textbf{Qualitative Analysis:}
Fig.~\ref{fig:attn_maps} shows representative attention maps from the hold-out test set. The model consistently highlights cellular regions with diagnostic value while suppressing background artifacts. Healthy slides display diffuse and low attention, reflecting the absence of dysplastic clusters. OPMD samples show focused activation around epithelial groups with mild nuclear atypia. OSCC images exhibit sharp, high-intensity attention on malignant clusters with pronounced pleomorphism. These patterns follow expected cytological cues and indicate that the \texttt{RAA}--MIL framework learns clinically meaningful features despite using only weak patient-level labels.

%\noindent\textbf{Summary of Findings:} \texttt{RAA}-MIL delivers consistent improvements across cross-validation andtest evaluations. The gains are pronounced for \textit{OPMD} and \textit{OSCC}, the clinically critical classes. The results show that region-level affinity enhances token representations and yields more stable MIL attention under weak patient-level supervision. These findings highlight the potential of region-aware token refinement for scalable, dataset-efficient cytology screening.

\section{Conclusion}

This study presents the first weakly supervised patient-level framework for oral cytology classification using region-aware MIL, demonstrating that \texttt{RAA} substantially improves the discriminative power of ViT-derived token representations under limited and heterogeneous clinical data. The proposed \texttt{RAA}-MIL achieves consistent gains in accuracy, weighted F1, and PR-AUC, particularly for OPMD and OSCC—two classes that are clinically most critical for early intervention. Despite these advances, the model remains constrained by the small size of the benign cohort, variability in staining and acquisition conditions, and the absence of multi-centre training data, which collectively limit its generalizability. Future work will expand the dataset across institutions, integrate stain-normalization and morphology-aware augmentations, and explore hierarchical MIL, multimodal fusion with patient metadata, and test-time adaptation strategies. Together, these directions can support a robust, scalable, and clinically deployable cytology AI system for early oral cancer detection.

% \section{Acknowledgments}
% \label{sec:acknowledgments}

\bibliographystyle{IEEEbib}
\bibliography{refs}

\end{document}